\documentclass[twocolumn,showpacs,aps,superscriptaddress,preprintnumbers,amsmath]{revtex4}
\usepackage{graphicx}
\usepackage{epsfig}
\usepackage{nicefrac}
\usepackage{placeins}
\newcommand{\BABARPubNumber}  {13/002}
\newcommand{\SLACPubNumber}  {15411}

\def\Btag	{\ensuremath{B_{\rm tag}}\xspace}  
\def\Bsig       {\ensuremath{B_{\rm sig}}\xspace}
\def\mes        {\ensuremath{m_{\rm ES}}\xspace}
\def\Eextra     {\ensuremath{E_{\rm extra}}\xspace}
\def\sB         {\ensuremath{s_B}\xspace}
\def\Npeak	{\ensuremath{N_{i}^{\rm peak}}\xspace}  
\def\Ncomb	{\ensuremath{N_{i}^{\rm comb}}\xspace}  
\def\Nbkg	{\ensuremath{N_{i}^{\rm bkg}}\xspace}  
\def\Nobs	{\ensuremath{N_{i}^{\rm obs}}\xspace}   
\def\eff	{\ensuremath{\eps_{i}^{\rm sig}}\xspace} 
\def\KorKstar   {\ensuremath{K^{(*)}}}
\def\knunu      {\ensuremath{\B\to \KorKstar\nu\nub}\xspace }
\def\kxnunu     {\ensuremath{\B\to K\nu\nub}\xspace }
\def\kpnunu     {\ensuremath{\Bp\to\Kp\nu\nub}\xspace }
\def\kznunu     {\ensuremath{\Bz\to\Kz\nu\nub}\xspace }
\def\ksnunu     {\ensuremath{\B\to\Kstar\nu\nub}\xspace }
\def\kspnunu    {\ensuremath{\Bp\to\Kstarp\nu\nub}\xspace }
\def\ksznunu    {\ensuremath{\Bz\to\Kstarz\nu\nub}\xspace }
\def\taunu    {\ensuremath{\Bp\to\taup\nut}\xspace }
\def\jnunu      {\ensuremath{\jpsi\to\nunub}\xspace}
\def\psinunu    {\ensuremath{\psitwos\to\nunub}\xspace}
\def\ccnunu    {\ensuremath{\ccbar\to\nunub}\xspace}
\input babarsym

\begin{document}

\begin{flushleft}
\babar-PUB-\BABARPubNumber \\
SLAC-PUB-\SLACPubNumber \\

\end{flushleft}

\title{\large \bf \boldmath Search for \knunu and invisible quarkonium decays}
%
\author{J.~P.~Lees}
\author{V.~Poireau}
\author{V.~Tisserand}
\affiliation{Laboratoire d'Annecy-le-Vieux de Physique des Particules (LAPP), Universit\'e de Savoie, CNRS/IN2P3,  F-74941 Annecy-Le-Vieux, France}
\author{E.~Grauges}
\affiliation{Universitat de Barcelona, Facultat de Fisica, Departament ECM, E-08028 Barcelona, Spain }
\author{A.~Palano$^{ab}$ }
\affiliation{INFN Sezione di Bari$^{a}$; Dipartimento di Fisica, Universit\`a di Bari$^{b}$, I-70126 Bari, Italy }
\author{G.~Eigen}
\author{B.~Stugu}
\affiliation{University of Bergen, Institute of Physics, N-5007 Bergen, Norway }
\author{D.~N.~Brown}
\author{L.~T.~Kerth}
\author{Yu.~G.~Kolomensky}
\author{M.~Lee
}
\author{G.~Lynch}
\affiliation{Lawrence Berkeley National Laboratory and University of California, Berkeley, California 94720, USA }
% ...
\collaboration{The \babar\ Collaboration}
\noaffiliation

\begin{abstract}
We search for the flavor-changing neutral-current decays \knunu, and the invisible decays \jnunu and \psinunu via $B\to \KorKstar\jpsi$ and $B\to \KorKstar\psitwos$ respectively, using a data sample of $471\times10^{6}$ \BB pairs collected by the \babar\ experiment. We fully reconstruct the hadronic decay of one of the $B$ mesons in the $\FourS\to\BB$ decay, and search for the \knunu decay in the rest of the event.  We observe no significant excess of signal decays over background and report branching fraction upper limits of $\BR(\kpnunu)<3.7\times10^{-5}$, $\BR(\kznunu)< 8.1\times10^{-5}$, $\BR(\kspnunu)<11.6\times10^{-5}$, $\BR(\ksznunu)<9.3\times10^{-5}$, and combined upper limits of $\BR(\kxnunu)<3.2\times10^{-5}$ and $\BR(\ksnunu)<7.9\times10^{-5}$, all at the 90\% confidence level. 
For the invisible quarkonium decays, we report branching fraction upper limits of $\BR(\jnunu)<3.9\times 10^{-3}$ and $\BR(\psinunu)<15.5\times 10^{-3}$ at the 90\% confidence level.  Using the improved kinematic resolution achieved from hadronic reconstruction, we also provide partial branching fraction limits for the \knunu decays over the full kinematic spectrum.  
\end{abstract}

\maketitle

\section{Introduction}
\label{sec:Introduction}

\begin{figure}
 \includegraphics[width=3.4in]{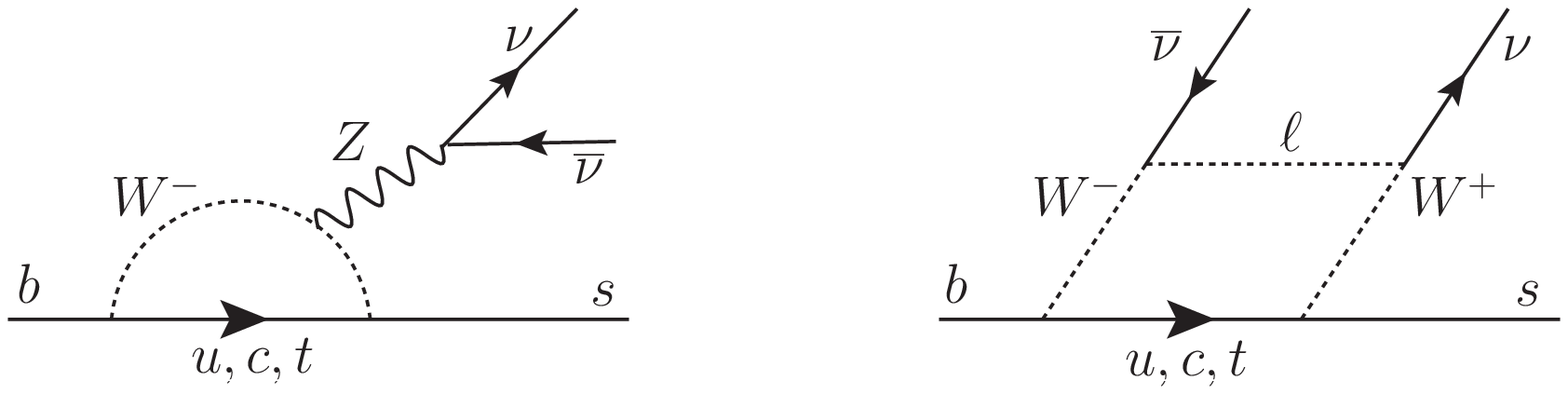}
 \caption{Lowest-order SM Feynman diagrams for $b\to s\nu\nub$ transitions. The virtual top quark provides the dominant contribution in each case.\label{fig:feynman}} 
\end{figure}

\begin{figure}
 \includegraphics[width=3.4in]{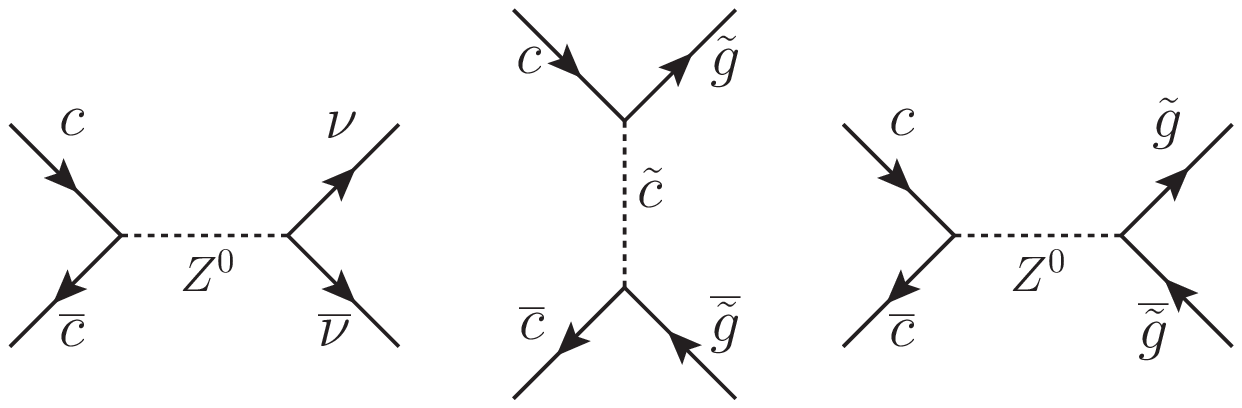} 
 \caption{Lowest-order Feynman diagrams of (from left to right) the SM decay \ccbar\to\nunub, the SUSY decay \ccbar into a pair of goldstinos ($\tilde{g}$) via a $c$-squark in the $t$-channel, and the SUSY decay \ccbar\to$\tilde{g}\overline{\tilde{g}}$ via a virtual \Z in the $s$-channel. \label{fig:feynman2}} 
\end{figure}

Flavor-changing neutral-current transitions, such as $b\to s\nu\nub$, are prohibited in the standard model (SM) at tree-level.  However, they can occur via one-loop box or electroweak penguin diagrams, as shown in Fig.~\ref{fig:feynman}. They can occur also in the SM via a quarkonium resonance state $b\to s\ccbar$, \ccnunu, where the \ccbar decay is mediated by a virtual \Z boson (Fig.~\ref{fig:feynman2}).  This latter decay process has the same final state as $b\to s\nu\nub$ with an additional constraint from the on-shell \ccbar mass.  Both the $b\to s\nu\nub$ and \ccnunu decay rates are expected to be small within the SM, with branching fractions estimated to be $\BR(\kpnunu) = \BR(\kznunu) = (4.5\pm 0.7)\times 10^{-6}$, $\BR(\kspnunu)= \BR(\ksznunu) = (6.8^{+1.0}_{-1.1})\times 10^{-6}$ \cite{ref:altmann}, and $\BR(\jnunu) = (4.54\times 10^{-7})\cdot\BR(\jpsi\to\epem)$ \cite{ref:chang}.  The $b\to s\nu\nub$ rates are predicted with smaller theoretical uncertainties than those in the corresponding $b\to s\ellp\ellm$ modes due to the absence of long-distance hadronic effects from electromagnetic penguin contributions.

Various new-physics scenarios exist that could significantly enhance the $b\to s\nu\nub$ branching fractions, as well as modify the expected SM decay distributions of $\sB\equiv q^2/m_B^2$, where $q^2$ is the squared magnitude of the four-momentum transferred from the $B$ meson to the neutrino pair, and $m_B$ is the $B$ meson mass.  Some of these scenarios predict massive particles that could contribute additional loop diagrams with similar amplitudes as those in the SM, such as nonstandard \Z couplings with supersymmetric (SUSY) particles \cite{ref:altmann}, fourth-generation quarks \cite{ref:buchalla}, anomalous top-charm transitions \cite{ref:LYY}, or a massive U(1) gauge boson $Z^{\prime}$ \cite{ref:altmann, ref:leptoZ}.  Since $b\to s\nu\nub$ has two final-state neutrinos, other sources of new physics can also contribute to the experimental signature of a kaon and missing four-momentum, such as low-mass dark-matter (LDM) candidates \cite{ref:bird, ref:mckeen, ref:darkBosons, ref:altmann}, unparticles \cite{ref:unparticle}, right-handed neutrinos \cite{ref:leptoZ}, or SUSY particles \cite{ref:FCNCportals}.  Models with a single universal extra dimension also predict higher decay rates \cite{ref:ColangeloKnunu}. 

The decays $\jnunu$ and $\psinunu$ provide additional windows for new-physics searches. In spontaneously-broken SUSY, a $\ccbar$ resonance can decay into a pair of goldstinos via either a virtual \Z in the $s$-channel or a $c$-squark exchange in the $t$-channel \cite{ref:chang} (Fig.~\ref{fig:feynman2}).  The contribution of a massive SU(2) gauge boson $Z^{\prime}$, introduced in the left-right SUSY model, could suppress the decay rates up to an order of magnitude \cite{ref:chang}.  Conversely, a low-mass U(1) gauge boson $U$ could enhance the invisible decay rates of quarkonium states by several orders of magnitude by coupling to LDM particles \cite{ref:McElrath, ref:fayet}.  The $U$ boson could decay into a pair of spin-\nicefrac{1}{2} Majorana ($\chi\chi$), spin-\nicefrac{1}{2} Dirac ($\chi\overline{\chi}$), or spin-0 ($\varphi\varphi$) LDM particles.

We  search for \kxnunu and \ksnunu, and for \jnunu and \psinunu via $B\to \KorKstar\jpsi$ and $B\to \KorKstar\psitwos$ respectively, where $\KorKstar$ signifies a charged or neutral \kaon or \Kstar meson \cite{ref:ccNote}.  We use a technique in which one $B$ meson is exclusively reconstructed in a hadronic final state before looking for a signal decay within the rest of the event. Since the four-momentum of one $B$ meson is fully determined, the missing mass resolution on the two final-state neutrinos and the suppression of background are improved with respect to other reconstruction techniques. 

Several previous searches for \kxnunu and \ksnunu have been performed by both the \babar\ and BELLE collaborations \cite{ref:elisaKnunu, ref:carlKnunu, ref:newBelle, ref:belleKnunu, ref:jackKnunu}.  Currently, the most stringent published upper limits at 90\% confidence level (CL) are $\BR(\kpnunu)<1.3\times 10^{-5}$ \cite{ref:carlKnunu} and $\BR(\ksnunu)<8\times 10^{-5}$ \cite{ref:elisaKnunu}. The $\BR(\kpnunu)$ limit was determined using semileptonic-tag reconstruction, which produces samples that are statistically larger and independent of those produced using the hadronic-tag reconstruction employed in this search. The $\BR(\ksnunu)$ limit was a combination of two \babar\ analyses, one using semileptonic-tag reconstruction and the other using hadronic-tag reconstruction.

A \jnunu search via $\psitwos\to\pip\pim\jpsi$ was performed by the BES collaboration, which set an upper limit at 90\% CL of $\BR(\jnunu) <1.2\times 10^{-2}\cdot \BR(\jpsi\to\mu^+\mu^-)$ \cite{ref:BES}. This article presents the first search for \jnunu using the hadronic-tag reconstruction of a $B$ meson decay. A search for \psinunu has not been performed previously.

\section{\boldmath The \babar\ Detector and Data Sample}
\label{sec:babar}

This search uses a data sample of $471 \pm 3$ million \BB pairs, corresponding to an integrated luminosity of $429\invfb$ collected at the \FourS resonance \cite{ref:lumi}.  The data were recorded with the \babar\ detector \cite{ref:babar} at the \pep2 asymmetric-energy $e^+e^-$ storage rings. The charged-particle tracking system consists of a five-layer double-sided silicon vertex tracker and a 40-layer drift chamber, both coaxial with a $1.5\,{\rm T}$ solenoidal magnetic field. Charged kaons and pions are distinguished by specific ionization energy-loss measurements from the tracking system for lower momentum particles, and by measurements from a ring-imaging Cherenkov radiation detector for higher momentum particles. A CsI(Tl) electromagnetic calorimeter is used to reconstruct photons of energy greater than 20\mev and to identify electrons.  Muon identification is provided by the instrumented flux return of the magnet.  Particle identification (PID) algorithms are trained to identify charged particle types by using 36 input parameters including momentum, polar and azimuthal angles, the Cherenkov angle, and energy-loss measurements \cite{ref:NIMU}.  We employ PID criteria that select $K^+$ mesons with an efficiency greater than 85\% and with approximately 1\% misidentification probability for pions and muons. 

Signal and background decays are studied using Monte Carlo (MC) samples simulated with Geant4 \cite{ref:geant4}.  The simulation includes a detailed model of the \babar\ detector geometry and response. Beam-related background and detector noise are extracted from data and are overlaid on the MC simulated events. Large MC samples of generic $\BB$ and continuum ($e^+e^-\to\tautau$ or $e^+e^-\to\qqbar$, where $q = u,d,s,c$) events provide ten times the number of $\FourS\to\BB$ and $e^+e^-\to\ccbar$ events as in the data sample, and four times the number of other continuum decays.  The $\FourS\to\BB$ signal MC samples are generated with one $B$ meson decaying via \knunu, with and without the \ccbar resonances, while the other $B$ meson decays according to a model tuned to world averages of allowed decay channels.  The \sB distributions for \knunu decays within signal MC samples are generated initially using a phase-space model, and then reweighted using the model from Ref.~\cite{ref:altmann}, henceforth referred to as ABSW.  Within \ksnunu decays, this model is also used to reweight the helicity-angle distribution between the signal $B$ and the $K^+$ or $K^0$ flight directions in the \Kstar rest frame.  The helicity amplitudes for the decay channels $B\to\Kstar\jpsi$ and $B\to\Kstar\psitwos$ are generated using values taken from a \babar\ measurement \cite{ref:jpsikamp}.

\section{\boldmath Analysis Method}
\label{sec:Analysis}

Event selection for both the \knunu and $B\to \KorKstar\ccbar$, $\ccnunu$ searches begins by fully reconstructing a $B$ meson (\Btag) in one of many hadronic final states, $\Bbar\to SX_{\rm had}^-$, where $S$ is a ``seed" meson ($D^{(*)+}$, $D^{(*)0}$, $D^{(*)+}_s$, or $J/\psi$) and $X_{\rm had}^-$ is a collection of at most five mesons, composed of charged and neutral kaons and pions with a net charge of $-1$.  This method, which was used also in Ref.~\cite{ref:Dtaunu}, reconstructs additional modes with respect to previous hadronic-tag \knunu analyses \cite{ref:elisaKnunu, ref:jackKnunu}, and results in approximately twice the reconstruction efficiency.  The $D$ seeds are reconstructed in the decay modes $D^+\to\KS\pip$, $\KS\pip\piz$, $\KS\pip\pip\pim$, $ K^-\pip\pip$, $K^-\pip\pip\piz$, $\Kp\Km\pip$, $\Kp\Km\pip\piz$; $\Dz\to K^-\pip$, $K^-\pip\piz$, $K^-\pip\pip\pim$, $\KS\pip\pim$, $\KS\pip\pim\piz$, $\Kp\Km$, $\pip\pim$, $\pip\pim\piz$, and $\KS\piz$. Additional seeds are reconstructed as $D^{*+}\to \Dz\pip$, $\Dp\piz$; $\Dstarz\to \Dz\piz$, $\Dz\g$; $D_s^{*+}\to D_s^+\g$; $D_s^+\to\phi[\to\Kp\Km]\pip$, $\KS\Kp$; and $J/\psi\to\epem$, $\mu^+\mu^-$. The \KS candidates are reconstructed via their decay to $\pip\pim$.

 Well-reconstructed \Btag candidates are selected using two kinematic variables: $\Delta E= E_{\Btag}-\sqrt{s}/2$ and $\mes = \sqrt{s/4 - {\vec p}^{\,2}_{\Btag}}$, where $E_{\Btag}$ and ${\vec p}_{\Btag}$ are the energy and momentum vector of the \Btag candidate, respectively, in the \epem center-of-mass (CM) frame and $\sqrt{s}$ is the total energy of the \epem system.  The value of $\Delta E$, which peaks at zero for correctly reconstructed $B$ mesons, is required to be between $-0.12$ and $0.12\gev$ or within two standard deviations around the mean for a given $X_{\rm had}^-$ mode, whichever is the tighter constraint.  If more than one \Btag candidate is reconstructed, the one in the mode with the highest purity (fraction of candidates that are correctly reconstructed within a given \Btag decay mode) is chosen.  If there are multiple candidates with the same purity, the one with the smallest $\lvert\Delta E\rvert$ is selected.

After requiring that the event contains between one and three charged tracks not used in the \Btag reconstruction (``signal-side" tracks), the purity of each mode is recalculated, and only the \Btag modes that have a recalculated purity greater than 68\% are retained.  This results in a total of 448 final states.  This purity value was optimized by maximizing the figure of merit \cite{ref:punzi} 
	\begin{equation}
	\label{punzi}
	\frac{\eff}{\tfrac{1}{2}n_{\sigma}+\sqrt{\Nbkg}} ,
	\end{equation}
where the number of sigmas $n_{\sigma} = 1.28$ corresponds to a one-sided Gaussian limit at 90\% CL, \eff is the total signal efficiency, and \Nbkg is the expected number of background events, with $i$ representing one of the signal decay channels. All other selection criteria discussed henceforth were optimized simultaneously using this same figure of merit.

\begin{figure}
 \includegraphics[width=3.2in]{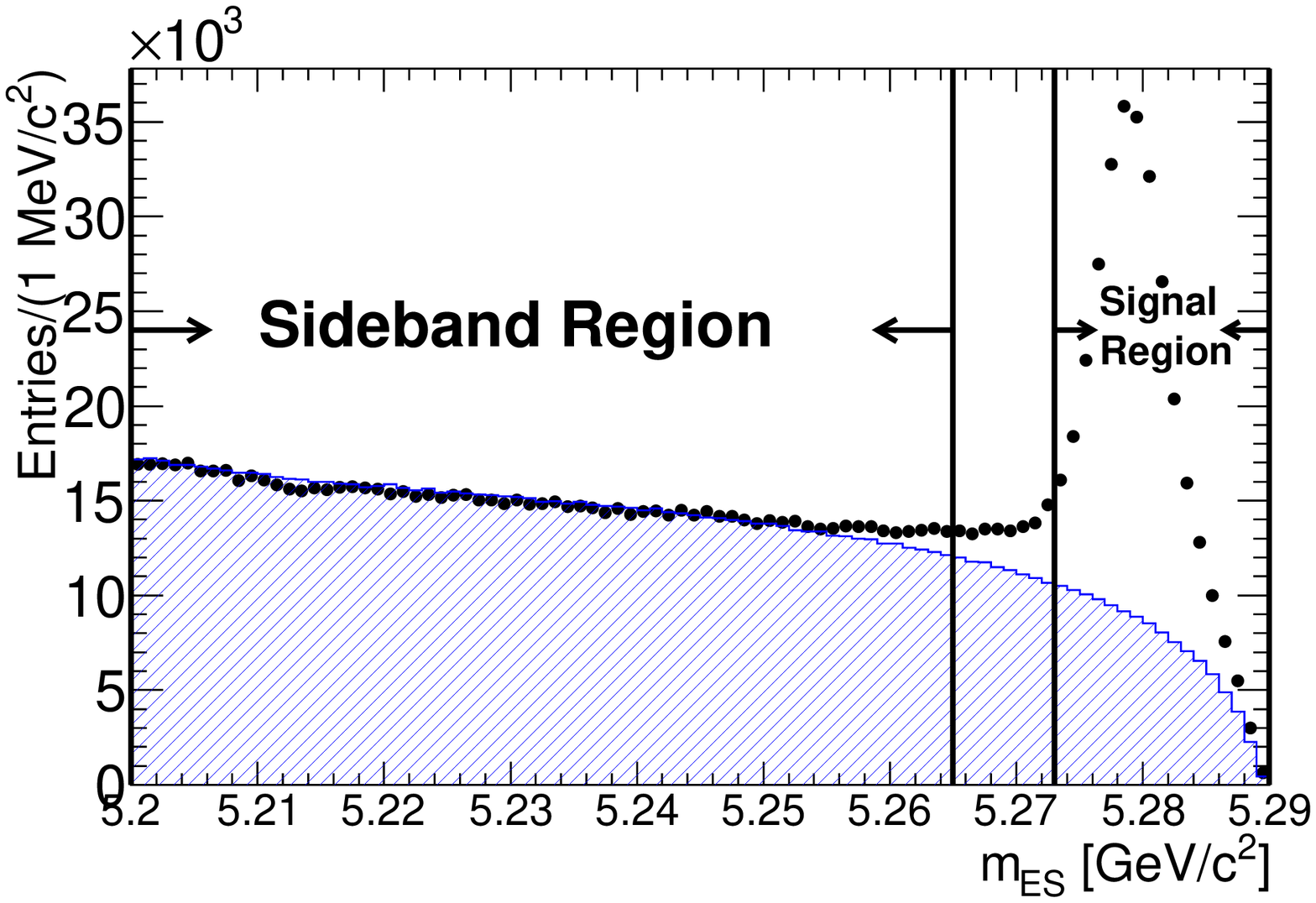}
 \caption{(color online) The \mes distribution for the \Btag candidates in data (points) and in the expected combinatorial background as predicted by the MC (shaded). This distribution includes only the charged and neutral \Btag candidates that pass the purity restrictions, the multivariate continuum suppression, and a requirement of one to three signal-side tracks.  The data within the \mes sideband region is used to extrapolate the expected number of combinatorial background events within the signal region.  \label{fig:mes}} 
\end{figure}

The signal region of the \Btag candidate is defined as $5.273<\mes<5.290\gevcc$ (Fig.~\ref{fig:mes}), since correctly reconstructed $B$ mesons produce a peak in this region near the nominal $B$-meson mass.  The \Btag candidates that are incorrectly reconstructed (``combinatorial" events), which result from continuum events or are due to particles assigned to the wrong $B$ meson, produce a distribution that is relatively uniform below the \mes signal region and decreases toward the kinematic limit within it.  Approximately 0.3\% of signal MC events and 12.0 million data events contain a \Btag that is reconstructed using the above requirements and found to be within the \mes signal region.

Since $B$ mesons are spin zero and are produced with low momentum in the CM frame ($\sim\!0.32\gevc$), their decay products are more isotropically distributed than non-\BB background.  For example, $\lvert\cos\theta_{\rm T}\rvert$, where $\theta_{\rm T}$ is the angle in the CM frame between the \Btag thrust \cite{ref:thrust} axis and the thrust axis of all other particles in the event, has a uniform distribution for \BB events but peaks near one for continuum events.  Continuum background is suppressed by using a multivariate likelihood selector based on six event-shape variables.  These consist of $\lvert\cos\theta_{\rm T}\rvert$, the cosine of the angle between ${\vec p}_{\Btag}$ and the beam axis, the magnitude of the \Btag thrust, the component of the \Btag thrust along the beam axis, the angle between the missing momentum vector (${\vec p}_{\rm miss}$) and the beam axis, and the ratio of the second-to-zeroth Fox-Wolfram moment \cite{ref:foxWolf} computed using all charged and neutral particles in the event.  The multivariate selector requires
	\begin{equation}
	\label{contlike}
	 \calL_{B}\equiv \frac{\prod_j {\cal P}_{B}(x_j)}{\prod_j {\cal P}_{B}(x_j)+\prod_j {\cal P}_{q}(x_j)}>53\% ,
	\end{equation}
where ${\cal P}_{q}(x_j)$ and ${\cal P}_{B}(x_j)$ are probability density functions determined from MC that describe continuum and signal-like \BB events, respectively, for the six event-shape variables $x_j$.  The $\calL_{B}$ requirement, which was optimized with other selection criteria using Eq.~\eqref{punzi}, also improves the agreement between data and MC by suppressing unmodeled continuum backgrounds.

In the sample of selected \Btag candidates, signal events are chosen such that a single $\KorKstar$ candidate can be reconstructed within the rest of the event and no additional charged tracks remain in the event.  The sum of the $\KorKstar$ and \Btag candidate charges must equal zero. Since signal decays have two final-state neutrinos, these events are required to have missing energy greater than zero,  where the missing energy is defined as the CM energy minus all detected calorimeter deposits from charged and neutral particles in the event. For \knunu, the signal decays are reconstructed in six channels: \kpnunu; $\Bz\to K^0\nu\nub$ where $K^0\to\KS$; $\kspnunu$, where $\Kstarp\to\Kp\piz$ and  $\Kstarp\to\KS\pip$; and $\ksznunu$, where $\Kstarz\to\Kp\pim$ and $\Kstarz\to\KS\piz$. For \ccnunu, the same six signal channels are employed with an additional requirement that the $\KorKstar$ momentum is consistent with a two-body decay, either \B\to $\KorKstar\jpsi$ or \B\to $\KorKstar\psitwos$. The \jpsi and \psitwos mesons then decay into a pair of neutrinos, thus yielding the same final states as for \knunu. 

We reconstruct $\KS\to\pip\pim$ decay candidates using two tracks of opposite charge, which originate from a common vertex and produce an invariant mass within $\pm 7\mevcc$ of the nominal \KS mass \cite{ref:pdg}.  The PID for each track must be inconsistent with that for an electron, muon, or kaon. The \piz candidates are reconstructed from pairs of photon candidates with individual energies greater than 30\mev, a total CM energy greater than 200\mev, and a $\g\g$ invariant mass between 100 and 160\mevcc.  All \Kp candidates must satisfy the PID criteria for a kaon. 

Reconstructed \Kstar candidates are required to have an invariant mass within  $\pm70\mevcc$ of the nominal \Kstar mass \cite{ref:pdg}.  A $\Kstarp\to\KS\pip$ candidate combines a \KS candidate with a track that satisfies the PID criteria for a pion.  If more than one $\Kstarp\to\KS\pip$ candidate can be reconstructed in an event, the one with the mass closest to the nominal \Kstarp mass is chosen.   A $\Kstarz\to\Kp\pim$ candidate combines one track that satisfies the PID criteria for a kaon with one that is inconsistent with the PID criteria for an electron, muon, or kaon.  In an event containing a \Kp (\KS) candidate and no additional signal-side tracks, $\Kstar\to\Kp\piz$ ($\KS\piz$) candidates are reconstructed if the invariant mass of a \piz candidate and the \Kp (\KS) candidate falls within the \Kstar mass window; otherwise the event is considered for the \Kp (\KS) signal channel.  If more than one $\Kstarp\to\Kp\piz$ or $\Kstarz\to\KS\piz$ candidate can be reconstructed, the one with the highest energy \piz candidate is chosen.  

\begin{figure}[!tbp]
 	\includegraphics[width=2.8in]{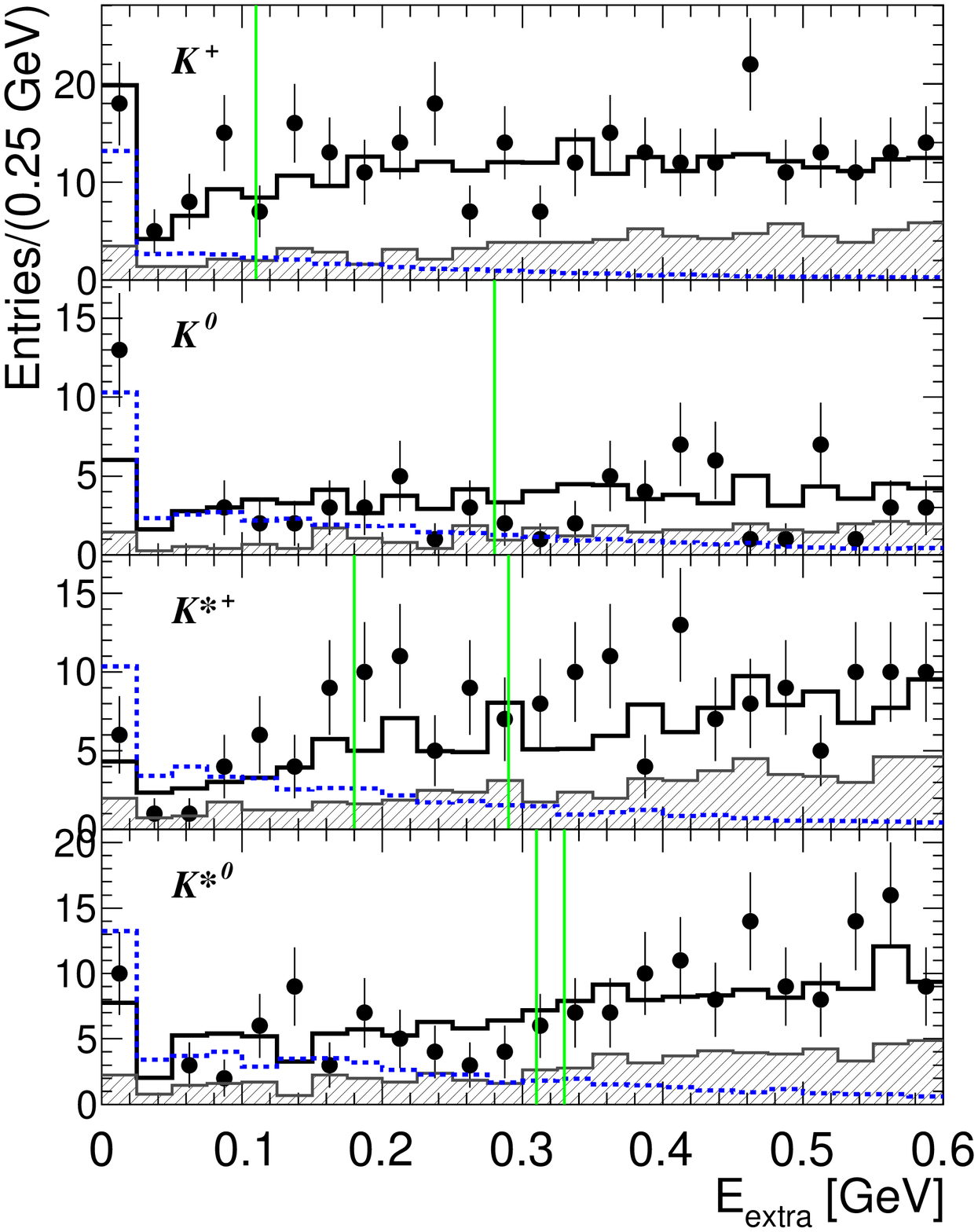} 
 	\caption{\label{fig:Eextra} (color online) The \Eextra distribution over the full \sB spectrum in the (from top to bottom) \Kp, \Kz, \Kstarp, and \Kstarz channels  after applying all other signal selection criteria.  The expected combinatorial (shaded) plus \mes-peaking (solid) background contributions are overlaid on the data (points).  The \knunu signal MC distributions (dashed) have arbitrary normalization. Both the \knunu and \ccnunu searches select events to the left of the vertical line that corresponds to the $E_i$ value of that channel, as given in Table \ref{tab:Eextra}.}
 \end{figure}

Once the \Btag and $\KorKstar$ are identified, the signal events are expected to contain little or no additional energy within the calorimeter.  However, additional energy deposits can result from beam-related photons, hadronic shower fragments that were not reconstructed into the primary particle deposit, and photons from unreconstructed $\Dstar\to D\g/\piz$ transitions in the \Btag candidate.  Only deposits with energy greater than $50\mev$ in the rest frame of the detector are considered, and the sum of all such additional energy deposits (\Eextra) is required to be less than a threshold value ($E_i$).  The values of $E_i$, given in Table \ref{tab:Eextra} and depicted in Fig.~\ref{fig:Eextra}, were optimized with the other selection criteria but were allowed to differ between signal channels. For events within the \Kp signal channel, calorimeter deposits identified as kaon shower fragments are not included in the \Eextra sum.  A fragment candidate is defined as a neutral calorimeter deposit whose momentum vector, when compared to that of the signal track, is separated by polar and azimuthal angles (relative to the beam axis and in the rest frame of the detector) of $\Delta\theta$ and $\Delta\phi$, respectively, such that $r_{\rm clus}<15^{\circ}$, where $r_{\rm clus} \equiv \sqrt{(\Delta\theta)^2 + \frac{2}{3}(Q_K\cdot \Delta\phi - 8^{\circ})^2}$ and $Q_K=\pm1$ is the $K^\pm$ charge.  The $r_{\rm clus}$ and fragment candidate definitions were optimized using studies of truth information in the signal MC samples.  The recovery of these kaon shower fragments improves the final signal efficiency in the $\Kp$ channel by about 13\%.  This procedure was explored for the other signal tracks, but the effect was small.

\begin{table}[!htp] 
    \begin{center}    
    \caption{\label{tab:Eextra} Threshold values $E_i$ for the \Eextra variable in each of the signal channels, determined using Eq.~\eqref{punzi}.  The channels in brackets refer to the \Kstar decay products.}
    \begin{ruledtabular}
    \begin{tabular}{lcccccc}   
        Channel & \Kp  & $K^0$ & $[K^+\piz]$ & $[\KS\pip]$ &  $[K^+\pim]$ &  $[\KS\piz]$   \\ \hline
        $E_i$ [GeV] & 0.11 & 0.28 & 0.18 & 0.29  & 0.31  & 0.33 \\   
    \end{tabular} 
    \end{ruledtabular}
    \end{center}   
\end{table}

The searches for \knunu and for \ccnunu via $B\to \KorKstar\ccbar$ diverge in the final step of the signal selection, which involves restricting the kinematics of the decay.  The value of \sB is calculated as $(p_{\Bsig}-p_{\KorKstar})^2/m_B^2$, where $p_{\KorKstar}$ is the four-momentum of the $\KorKstar$ candidate, and $p_{\Bsig}$ is the expected signal $B$ four-momentum with an energy of $\sqrt{s}/2$, the nominal $B$-meson mass, and a momentum vector pointing opposite the \Btag momentum.  For \knunu, the signal region optimized for maximum SM sensitivity is $0<\sB<0.3$ for all six signal channels.  This corresponds to a $\KorKstar$ momentum greater than about 1.8 (1.7)\gevc in the signal $B$ rest frame for \kxnunu (\ksnunu) events.  Partial branching fractions over the full \sB spectrum are also provided for sensitivity to new-physics scenarios that modify the kinematic distributions for \knunu.  For \ccnunu via $B\to \KorKstar\ccbar$, the invariant mass of the two neutrinos $m_{\nunub} \equiv \sqrt{\sB m_B^2}$ is expected to correspond to the mass of the \jpsi (3.097 $\gevcc$) meson or to that of the \psitwos (3.686 $\gevcc$) meson.  Signal events are selected within three standard deviations around the nominal \ccbar masses, which results in windows of $3.044 < m_{\nu\nub} <3.146$ ($3.019 < m_{\nunub}<3.175$)\gevcc for the $B\to\kaon\jpsi$ ($B\to\Kstar\jpsi$) channels, and $3.650 < m_{\nu\nub} <3.724$ ($3.627 < m_{\nunub}<3.739$)\gevcc for the $B\to\kaon\psitwos$ ($B\to\Kstar\psitwos$) channels.

To avoid experimenter bias, all the above selection criteria and values were optimized using the MC before looking at any data events within the \Eextra and \mes signal regions.

\section{\boldmath Background and Branching Fraction Extraction }
\label{sec:Bkg}

The total number of background events \Nbkg in the signal region has two components: \Npeak is the number of expected background events having a correctly reconstructed \Btag candidate and hence peaking within the \mes signal region, and \Ncomb is the number of expected combinatorial background events, including both continuum events and \BB events with an incorrectly reconstructed \Btag candidate.  To reduce the dependence on MC simulations, the number of \Ncomb events is extrapolated directly from the observed data events within the \mes sideband region, defined as $5.200<\mes<5.265\gevcc$ and depicted in Fig.~\ref{fig:mes}.  The shape of the combinatorial \mes distribution is estimated using MC samples of continuum events and of \BB events reconstructed with the wrong charge.

The number of \Npeak events is estimated from generic \BB MC samples. Over half of \Npeak is found to be from $B\to D^{(*)}\ell\nu$ ($\ell= e$ or $\mu$) decays in which no lepton candidate is identified in the event and the $\KorKstar$ is a daughter of the $D$ or \Dstar meson.  One particular peaking background in the \knunu search is \taunu, with $\taup\to K^{(*)+}\nub_{\tau}$, which has the same final state as the signal decay \cite{ref:bartsch}. Exclusive \taunu MC samples, assuming a branching fraction of ($1.65\pm 0.34)\times 10^{-4}$ \cite{ref:pdg}, indicate that this background constitutes less than 15 (5)\% of the total background in the \kpnunu (\kspnunu) channel. 

Since both \Npeak and \eff are determined from MC samples, we normalize the MC yields to the data to account for differences between data and MC, such as from the \Btag reconstruction and the modeled branching fractions of \Btag modes within the MC.  This normalization is performed before applying the full signal selection in order to have a large background-to-signal ratio; looser $\KorKstar$ mass windows and \Eextra selection requirements are used such that the number of background events is approximately 60 times larger than the final background contribution, over the full \sB spectrum.  The peaking background component in the \BB MC is then normalized to the number of data events that peak within the \mes signal region.  This peaking yield normalization is performed separately for charged and neutral \Btag candidates, and results in the scaling of all signal and background MC samples by $1.027\pm0.039$ $(1.017\pm0.044)$ for charged (neutral) \Btag candidates.

The signal branching fractions are calculated using
\begin{equation}
        \label{BFeq}
        \BR_{i} = \frac{\Nobs-(\Npeak+\Ncomb)}{\eff N_{\BB}},\end{equation}
 where $N_{\BB}=471\times10^{6}$ is the total number of $B$ meson pairs in the data sample and \Nobs is the number of data events within the signal region.  The total signal efficiency \eff includes that of the \Btag reconstruction and is determined separately for each of the signal channels $i$.  Since misreconstructed events from other signal channels contribute to \Npeak, the branching fractions of all signal channels are determined simultaneously by inverting a $6\times6$ efficiency matrix $\eps_{ij}$, which describes the probability that a signal event of process $i$ is reconstructed in signal channel $j$. Branching fraction limits and uncertainties are computed using a mixed frequentist-Bayesian approach described in Ref.~\cite{ref:barlow}, with the systematic uncertainties on \Nbkg and \eff modeled using Gaussian distributions. To combine the results of signal decay channels, we find the $\BR_{i}$ value that maximizes a likelihood function defined as the product of the Poisson probabilities of observing $N_{i}^{\rm obs}$ events.

\section{\boldmath Systematic Studies}
\label{sec:Systematics}

To verify the modeling of \eff and \Nbkg, a control sample of $B\to D\ell\nu$ events is selected.  In place of a signal $K^{*}$ candidate, the events are required to contain a reconstructed $\Dz\to\Km\pip$, $\Dm\to\Kp\pim\pim$, or $\Dm\to\KS\pim$ candidate with an invariant mass within $\pm 35\mevcc$ of the nominal $D$-meson mass values \cite{ref:pdg}.   The event must have one additional track that satisfies the PID criteria of either an electron or muon.  All other reconstruction and signal selection requirements are retained.  The resulting yields in the data agree with MC expectations, assuming the well-measured branching fractions of $B\to D\ell\nu$ \cite{ref:pdg}, within the 7\% (12\%) statistical uncertainty of the data in the $0<\sB<0.3$ (\jpsi or \psitwos mass) region.  

The control sample is used to determine the systematic uncertainties due to the MC modeling of the \Eextra variable within data. Additional uncertainties on \Npeak and \eff are due to the \KS and \Kstar mass reconstruction windows, the \piz reconstruction, and the uncertainties in the branching fractions \cite{ref:pdg} of the dominant backgrounds contributing to \Npeak.  The uncertainty on \Ncomb is dominated by the sideband data statistics.  Other systematic uncertainties, such as those from PID, tracking, \Btag reconstruction, $N_{\BB}$, and the assumption that charged and neutral \BB pairs are produced at equal rates, are all accounted for by the normalization of the MC peaking yields.  Because the peaking yield in data depends on the extrapolated shape of the combinatorial \Btag background, the normalization scale factors are re-evaluated by varying the method used to extrapolate this shape.  The resulting variations on the final \Nbkg and \eff values are taken as the systematic uncertainties due to the normalization.

Due to the approximately $1.0\%$ resolution on the \sB measurement around $\sB=0.3$, an uncertainty is evaluated within the \knunu signal region. Similarly, the resolution on $m_{\nunub}$ contributes to uncertainties within the \jnunu and \psinunu signal regions.  Only the systematic uncertainties due to the \Npeak branching fractions and to \sB or $m_{\nunub}$ differ between the \knunu, \jnunu, and \psinunu searches.  The systematic uncertainties are summarized in Tables \ref{tab:sysSumm} and \ref{tab:sysSum2}; the former lists the uncertainties shared by the searches, while the latter lists those that differ.

\begin{table}[!htbp]
	\begin{center}
  	\begin{ruledtabular}
	\caption{Summary of systematic uncertainties that are shared by the \knunu,  \jnunu, and \psinunu searches.  All values are relative uncertainties in \%.  The channels in brackets refer to the \Kstar decay products.} 
	\label{tab:sysSumm}
	\begin{tabular}{lcccccc} 
Source & {\footnotesize\Kp} & {\footnotesize$[K^+\piz]$} &{\footnotesize$[\KS\pip]$} & {\footnotesize$K^0$} & {\footnotesize$[K^+\pim]$}& {\footnotesize$[\KS\piz]$}  \\ \hline 
\eff normalization     & 3.5 	& 3.5 	 & 3.5 	 & 8.9   & 8.9 	 & 8.9 \\ 
\Nbkg normalization    & 2.3 	& 2.3 	 & 2.3 	 & 6.0   & 6.0 	 & 6.0 \\ 
\KS reconstruction     &  --    & --     & 1.4   & 1.4   & --    & 1.4   \\ 
\Kstar reconstruction  &  --    & 2.8    & 2.8   & --    & 2.8   & 2.8   \\ 
\piz reconstruction    &  --    & 3.0    & --    & --    & --    & 3.0   \\ 
\Eextra                &  4.5   & 6.0    & 6.5   & 6.0   & 6.0   & 6.5   \\ 

	\end{tabular}
   	\end{ruledtabular}
	\end{center}

	\begin{center}
  	\begin{ruledtabular}
	\caption{Summary of systematic uncertainties that differ between the \knunu, \jnunu, and \psinunu searches, and the total systematic uncertainties for each signal channel.  All values are relative uncertainties in \%.  The total systematic uncertainties are determined by adding in quadrature each relevant uncertainty, including those listed in Table \ref{tab:sysSumm}. }
	\label{tab:sysSum2}
	\begin{tabular}{lcccccc} 
Source & {\footnotesize\Kp} & {\footnotesize$[K^+\piz]$} &{\footnotesize$[\KS\pip]$} & {\footnotesize$K^0$} & {\footnotesize$[K^+\pim]$}& {\footnotesize$[\KS\piz]$}  \\ \hline
\hline
\multicolumn{7}{c}{\knunu}\\
\Npeak \,\,{\BR}\,'s & 2.8    & 2.8	 & 2.8	 & 2.8   & 2.8   & 2.8   \\ 
\sB resolution         &  3.6   & 3.6    & 3.6   & 3.6   & 3.6   & 3.6   \\  
Total \Npeak syst.  & 6.8  & 8.9    & 8.8   & 9.7   & 10.0  & 10.9   \\ 
Total \Ncomb syst.  & 2.3  & 2.3    & 2.3   & 6.0   & 6.0   & 6.0  \\ 
Total \eff syst.    & 6.7  & 8.8    & 8.8   & 11.4  & 11.7  & 12.4 \\ 
\hline
\multicolumn{7}{c}{\jnunu}\\
\Npeak \,\,{\BR}\,'s 	& 3.5    & 3.5	 & 3.5	 & 3.5  & 3.5   & 3.5   \\ 
$m_{\nunub}$ resolution & 1.1   & 2.1    & 0.4   & 0.7  & 0.3   & 1.3   \\  
Total \Npeak syst.  	&  6.2  & 8.6    & 8.4   & 9.3  & 9.6  & 10.5    \\ 
Total \Ncomb syst.  	& 2.3  & 2.3    & 2.3   & 6.0   & 6.0   & 6.0  \\ 
Total \eff syst.    	& 5.8  & 8.3    & 8.0   & 10.8   & 11.1  & 11.9 \\ 
\hline
\multicolumn{7}{c}{\psinunu}\\
\Npeak \,\,{\BR}\,'s 	& 2.8    & 2.8	 & 2.8	 & 2.8  & 2.8   & 2.8   \\ 
$m_{\nunub}$ resolution & 0.8   & 2.4    & 1.0   & 0.9   & 1.8   & 3.1   \\  
Total \Npeak syst.  	& 5.8  & 8.5    & 8.1   & 9.1  & 9.5   & 10.7 \\ 
Total \Ncomb syst.  	& 2.3  & 2.3    & 2.3   & 6.0   & 6.0   & 6.0  \\ 
Total \eff syst.	& 5.8  & 8.4    & 8.1   & 10.9   & 11.2  & 12.2   \\ 
	\end{tabular}
   	\end{ruledtabular}
	\end{center}
\end{table}

\section{\boldmath Results for \knunu}
\label{sec:Physics}

\begin{figure}[!tbp]
        \includegraphics[width=2.8in]{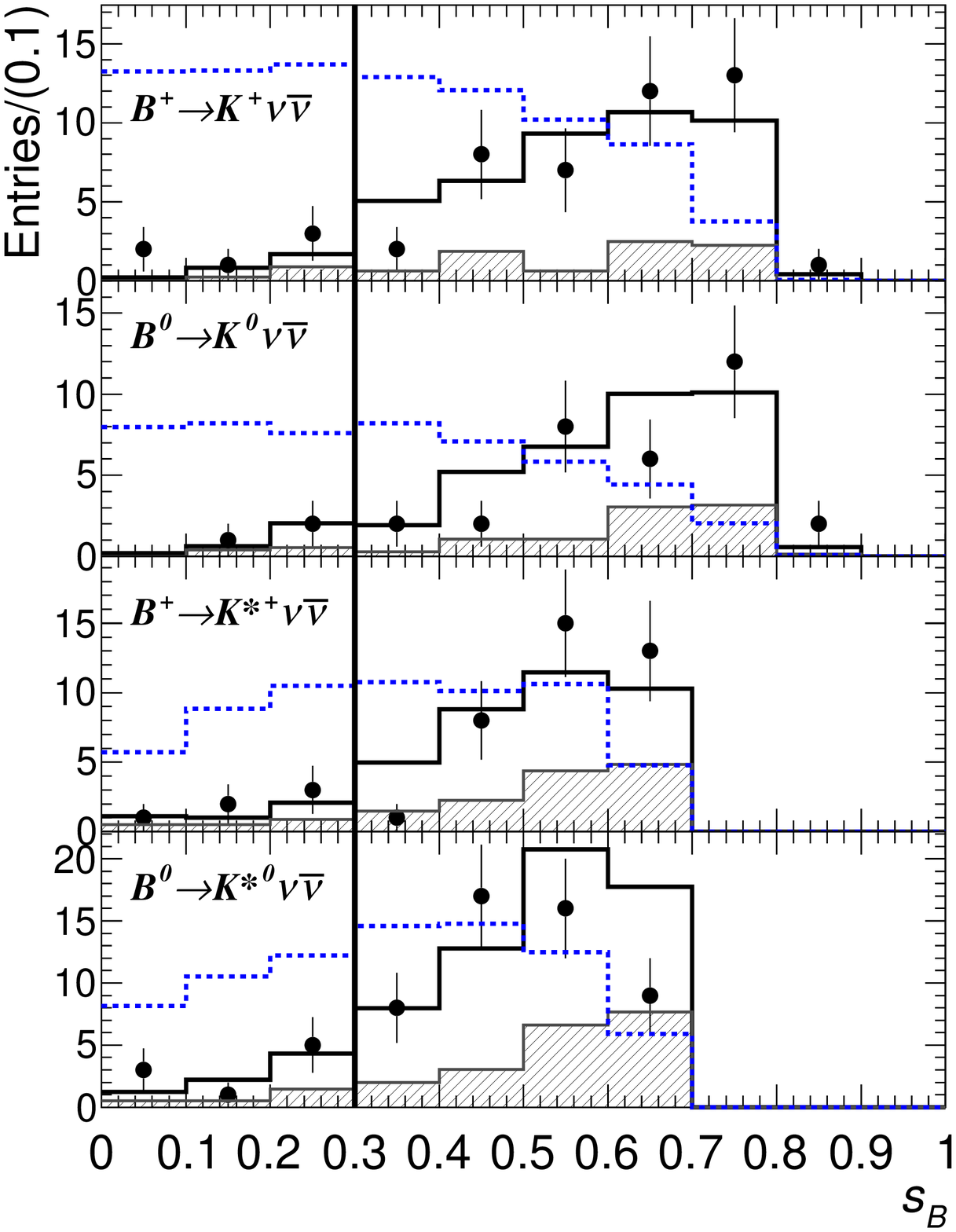}
        \caption{\label{fig:sBplots} (color online) The \sB distribution for (from top to bottom) \kpnunu, \kznunu, \kspnunu, and \ksznunu events after applying the full signal selection.  The expected combinatorial (shaded) plus \mes-peaking (solid) background contributions are overlaid on the data (points).  The signal MC distributions (dashed) are normalized to branching fractions of $20\times 10^{-5}$ for \kpnunu and $50\times 10^{-5}$ for the other channels.  Events to the left of the vertical lines are selected to obtain SM-sensitive limits, while the full spectra are used to determine partial branching fractions. }
 \end{figure}

Figure ~\ref{fig:sBplots} shows the observed data yields, expected background contributions, and SM signal distributions over the full \sB spectrum.  Tables \ref{tab:finalNums2} and \ref{tab:finalNums} summarize the number of observed data events within the \sB signal region ($0<\sB<0.3$), expected backgrounds, \knunu signal efficiencies, branching fraction central values, and branching fraction limits at the 90\% CL.  Combining the signal channels, we determine upper limits of $\BR(\kxnunu)<3.2\times10^{-5}$ and $\BR(\ksnunu)<7.9\times10^{-5}$.  Since we see a small excess over the expected background in the \Kp channel, we report a two-sided 90\% confidence interval.  However, the probability of observing such an excess within the signal region, given the uncertainty on the background, is 8.4\% which corresponds to a one-sided Gaussian significance of about $1.4\,\sigma$.  Therefore, this excess is not considered significant.

\begin{table*}[!tp]
	 \caption{Expected \ksnunu background yields $\Nbkg\!=\Npeak+\Ncomb$, signal efficiencies \eff, number of observed data events \Nobs, resulting branching fraction upper limits at 90\% CL, and the combined upper limits and central values, all within the $0<\sB<0.3$ region.  Uncertainties are statistical and systematic, respectively. The channels in brackets refer to the \Kstar decay products. \label{tab:finalNums2}}
	\begin{ruledtabular}
 	\begin{tabular}{lcccc}
 & $\Bp\to[\Kp\piz]\nu\nub$   & $\Bp\to[\KS\pi^+]\nu\nub$ &  $\Bz\to [K^+\pi^-]\nu\nub$ & $\Bz\to[\KS\piz]\nu\nub$ \\ \hline \hline

\Npeak     &   1.2 $\pm$ 0.4 $\pm$ 0.1 & 1.3 $\pm$ 0.4 $\pm$ 0.1 & 5.0 $\pm$ 0.8 $\pm$ 0.5 & 0.2 $\pm$ 0.2 $\pm$ 0.0  \\
\Ncomb     &   1.1 $\pm$ 0.4 $\pm$ 0.0 & 0.8 $\pm$ 0.3 $\pm$ 0.0 & 2.0 $\pm$ 0.5 $\pm$ 0.1 & 0.5 $\pm$ 0.3 $\pm$ 0.0  \\ 
\Nbkg      &   2.3 $\pm$ 0.5 $\pm$ 0.1 & 2.0 $\pm$ 0.5 $\pm$ 0.1 & 7.0 $\pm$ 0.9 $\pm$ 0.5 & 0.7 $\pm$ 0.3 $\pm$ 0.0  \\
\eff $(\times 10^{-5})$  &   4.9 $\pm$ 0.2 $\pm$ 0.4 & 6.0 $\pm$ 0.2 $\pm$ 0.5 & 12.2 $\pm$ 0.3 $\pm$ 1.4 & 1.2 $\pm$ 0.1 $\pm$ 0.1 \\
\Nobs     & 3 & 3 & 7 & 2 \\ 

Limit &  $<19.4\times10^{-5}$ & $<17.0\times10^{-5}$ & $<8.9\times10^{-5}$ & $<86\times10^{-5}$ \\  

$\BR(B^{+/0}\to K^{*+/0}\nu\nub)$ & \multicolumn{2}{c}{$(3.3_{-3.6}^{+6.2}$$_{-1.3}^{+1.7})\times10^{-5}$} & \multicolumn{2}{c}{$(2.0_{-4.3}^{+5.2}$$_{-1.7}^{+2.0})\times10^{-5}$} \\
Limit & \multicolumn{2}{c}{$<11.6\times10^{-5}$}  & \multicolumn{2}{c}{$<9.3\times10^{-5}$} \\ 

$\BR(\ksnunu)$ & \multicolumn{4}{c}{$(2.7_{-2.9}^{+3.8}$$_{-1.0}^{+1.2})\times10^{-5}$} \\
Limit & \multicolumn{4}{c}{$<7.9\times10^{-5}$} 

 	\end{tabular}
	\end{ruledtabular}
\end{table*}

\begin{table}[!tp]
	 \caption{Expected \kxnunu background yields $\Nbkg\!=\Npeak+\Ncomb$, signal efficiencies \eff, number of observed data events \Nobs, resulting branching fraction upper limits at 90\% CL, the central values $\BR_i$, and the combined upper limits and central value, all within the $0<\sB<0.3$ region.  Lower limits at 90\% CL are also reported, as discussed in the text.  Uncertainties are statistical and systematic, respectively. The \kznunu efficiency accounts for $\BR(K^0\to\KS)$ and $\BR(\KS\to\pip\pim)$ \cite{ref:pdg}.  \label{tab:finalNums}}
	\begin{ruledtabular}
 	\begin{tabular}{lcc}
           & \kpnunu                                     & \kznunu                             \\ \hline 
\Npeak     &  1.8 $\pm$ 0.4 $\pm$ 0.1 			&  2.0 $\pm$ 0.5 $\pm$ 0.2           \\ 
\Ncomb     &  1.1 $\pm$ 0.4 $\pm$ 0.0 			&  0.9 $\pm$ 0.4 $\pm$ 0.1         \\ 
\Nbkg      &  2.9 $\pm$ 0.6 $\pm$ 0.1  			&  2.9 $\pm$ 0.6 $\pm$ 0.2          \\
\eff $(\times 10^{-5})$   &  43.8 $\pm$ 0.7 $\pm$ 3.0 & 10.3 $\pm$ 0.2 $\pm$ 1.2 \\
\Nobs     & 6                             & 3                                    \\

$\BR_i$	&  $(1.5_{-0.8}^{+1.7}$$_{-0.2}^{+0.4})\times10^{-5}$ & $(0.14_{-1.9}^{+6.0}$$_{-0.9}^{+1.7})\times10^{-5}$\\ 
Limits & $(> 0.4, < 3.7)\times10^{-5}$ & $<8.1\times10^{-5}$ \\ 

$\BR(\kxnunu)$
	& \multicolumn{2}{c}{$(1.4_{-0.9}^{+1.4}$$_{-0.2}^{+0.3})\times10^{-5}$} \\ 
Limits & \multicolumn{2}{c}{$(> 0.2, < 3.2)\times10^{-5}$}  \\

 	\end{tabular}
	\end{ruledtabular}
\end{table}

Using the same procedure as when combining signal decay channels, the \kxnunu branching fraction central values are combined with a previous semileptonic-tag \babar\ analysis that searched within a statistically independent data sample \cite{ref:carlKnunu}. We obtain combined \babar\ upper limits at the 90\% CL of
\begin{equation}
	\label{SLtagCarl}
	\begin{split}
	\BR(\kpnunu)&< 1.6\times 10^{-5}, \\
 	\BR(\kznunu)&< 4.9\times 10^{-5},~{\rm and} \\
	\BR(\kxnunu)&< 1.7 \times 10^{-5}.
	\end{split}
\end{equation}
The combined central value is $\BR(\kxnunu)=(0.8^{+0.7}_{-0.6})\times 10^{-5}$, where the uncertainty includes both statistical and systematic uncertainties.    These combined results reweight the \sB distribution to that of the ABSW theoretical model (dashed curve in Fig.~\ref{fig:sBplots}), which decreases the signal efficiencies published in Ref.~\cite{ref:carlKnunu} by approximately $10\%$. The \ksnunu central values also can be combined with the semileptonic-tag results from a previous \babar\ search \cite{ref:elisaKnunu}.  In order to obtain approximate frequentist intervals, the likelihood functions in the previous search are extended to include possibly negative signals.  We obtain combined \babar\ upper limits at the 90\% CL of
\begin{equation}
	\label{SLtagElisa}
	\begin{split}
	 \BR(\kspnunu)&< 6.4\times 10^{-5}, \\
	 \BR(\ksznunu)&< 12\phantom{.}\times 10^{-5},~{\rm and} \\
         \BR(\ksnunu)&< 7.6\times 10^{-5}.
	\end{split}
\end{equation}
The combined central value is $\BR(\ksnunu)=(3.8^{+2.9}_{-2.6})\times 10^{-5}$.

\begin{figure}[!btp]
 	\includegraphics[width=1.63in]{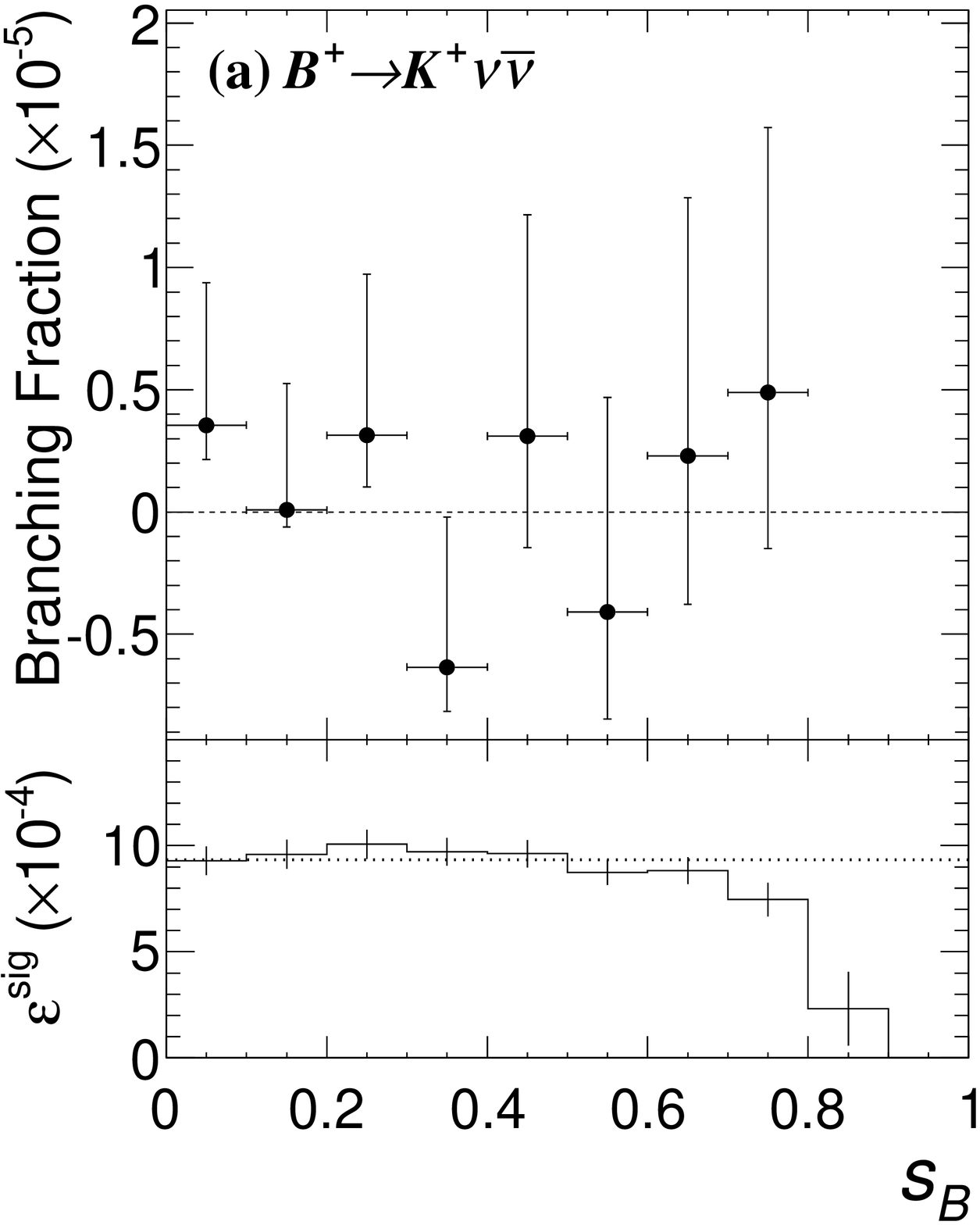}
	\hfill
 	\includegraphics[width=1.63in]{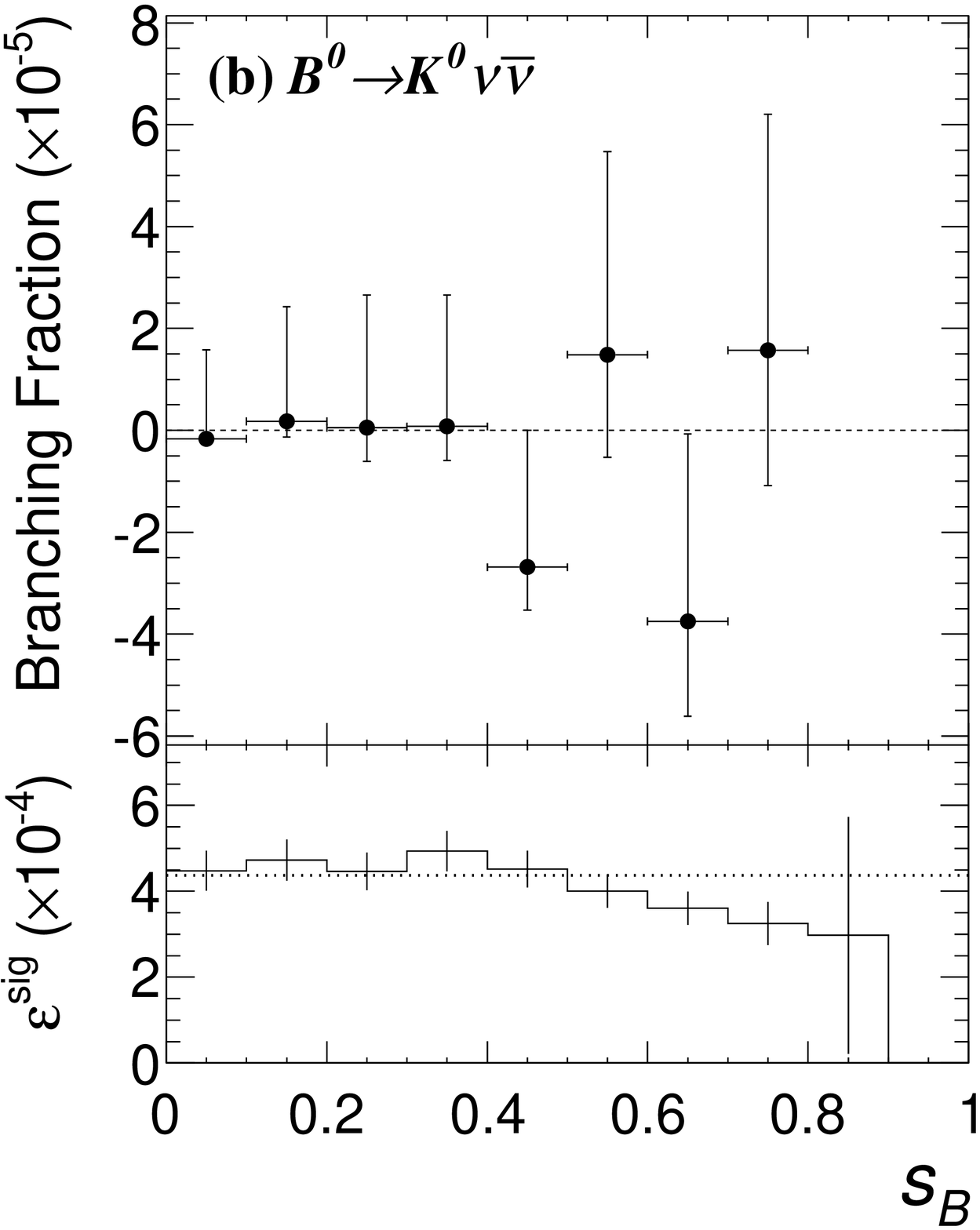} \\
 	\includegraphics[width=1.63in]{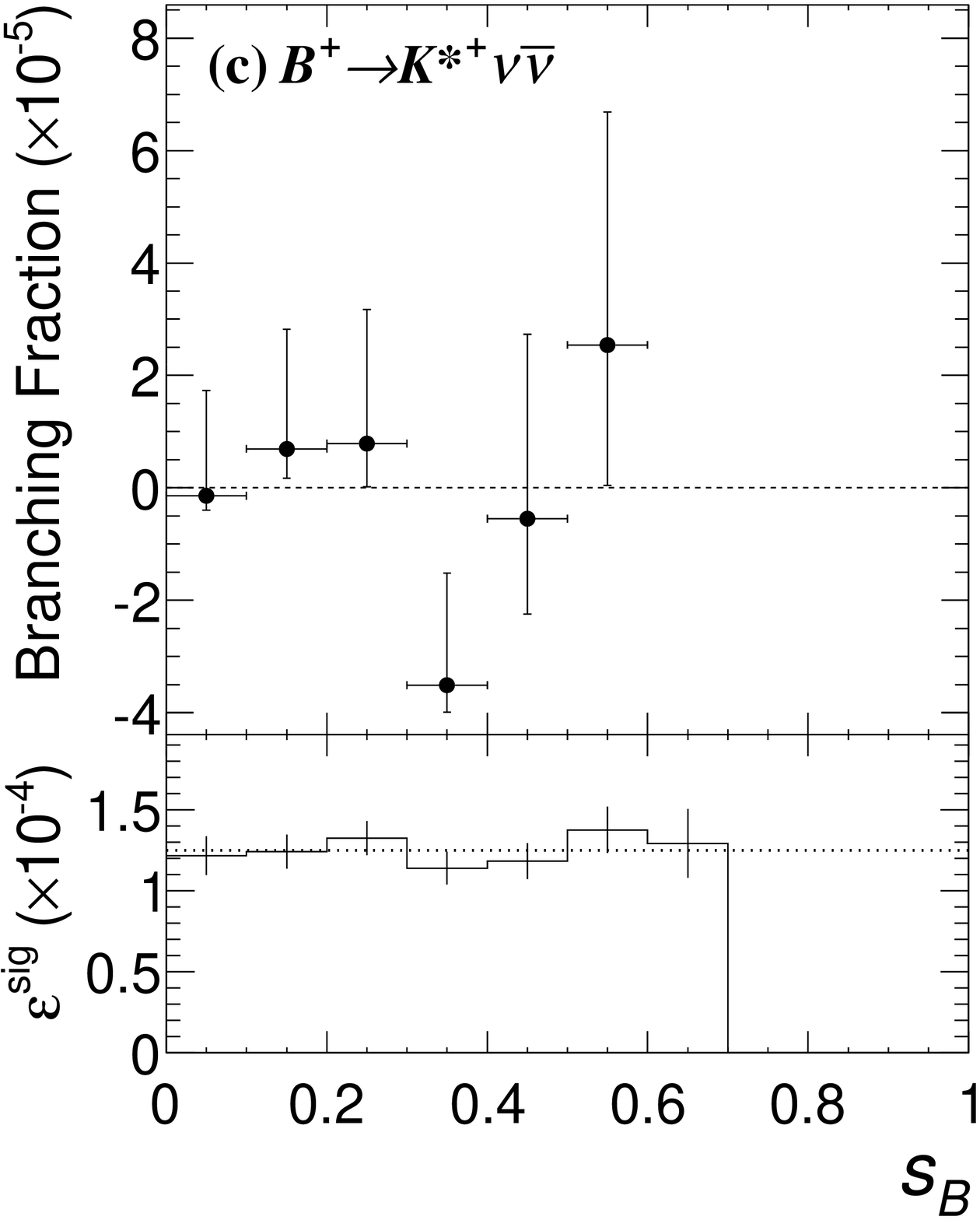} 
	\hfill
 	\includegraphics[width=1.63in]{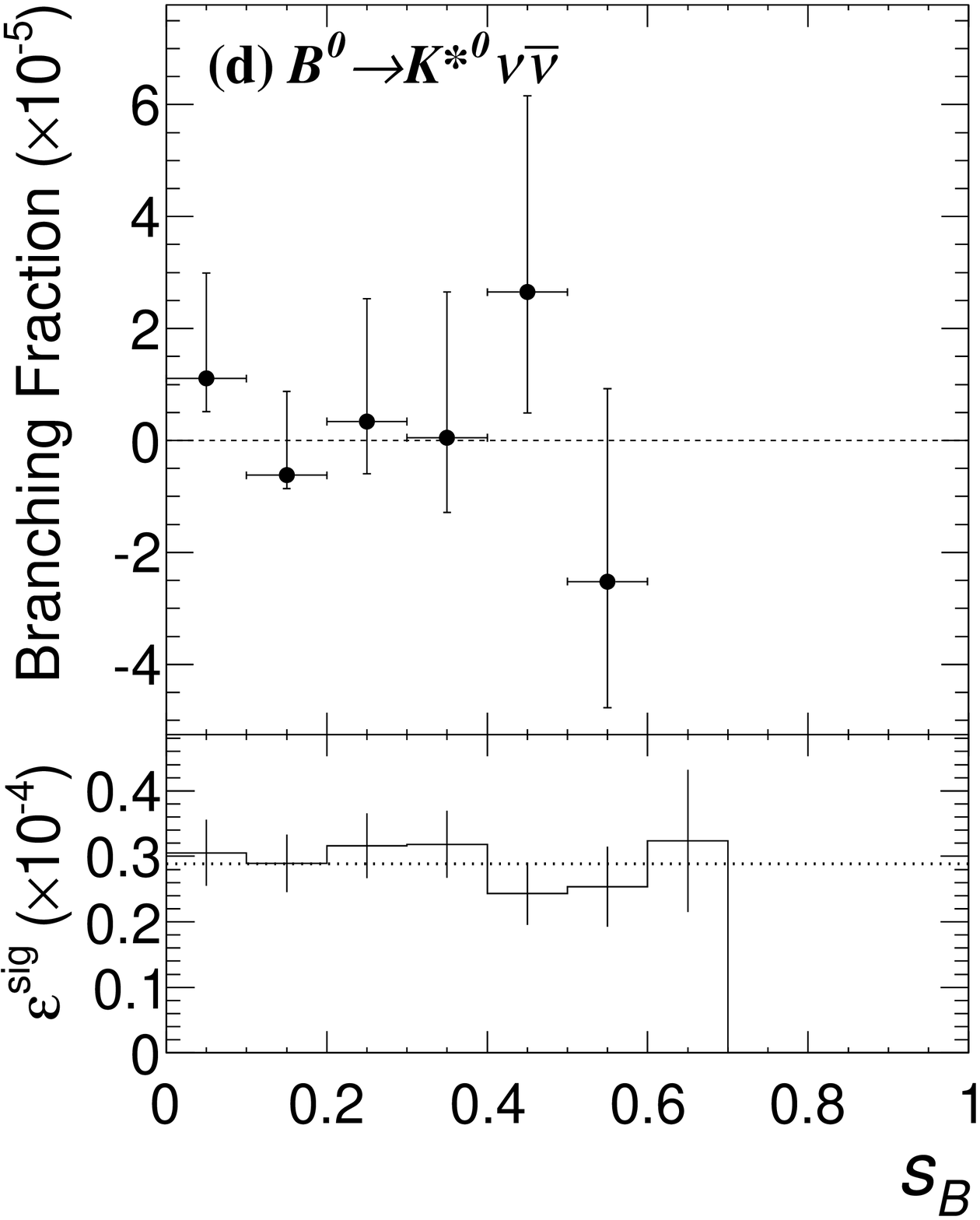} 
 	\caption{\label{fig:sBULs} The central values (points with $1\sigma$ error bars) of the partial branching fractions $\Delta\BR_i$ versus \sB, for (a) \kpnunu, (b) \kznunu, (c) \kspnunu, and (d) \ksznunu.  The subplots show the distribution of the final signal efficiencies within each \sB interval (histogram with error bars) and over the full \sB spectra (dotted line). The partial branching fractions are provided only within the intervals that are unaffected by the kinematic limit at large \sB.}
 \end{figure}

Since certain new-physics models suggest that enhancements are possible at high \sB values, we also report model-independent partial branching fractions ($\Delta\BR_i$) over the full \sB spectrum by removing the low-\sB requirement.  The $\Delta\BR_i$ values are calculated in intervals of $\sB=0.1$, using Eq.~\eqref{BFeq} (with the \Nobs, \Npeak, \Ncomb, and \eff values found within the given interval) multiplied by the fraction of the signal efficiency distribution inside that interval. Figure~\ref{fig:sBULs} shows the partial branching fractions.  The signal efficiency distributions are relatively independent of \sB, which are also illustrated in Fig.~\ref{fig:sBULs}. To compute model-specific values from these results, one can sum the central values within the model's dominant interval(s) (with uncertainties added in quadrature) and divide the sum by the fraction of the model's distribution that is expected to lie within the same \sB intervals.  These partial branching fractions provide branching fraction upper limits for several new-physics scenarios at the level of $10^{-5}$.

\begin{figure}[!btp]
        \center
        \includegraphics[width=3.0in]{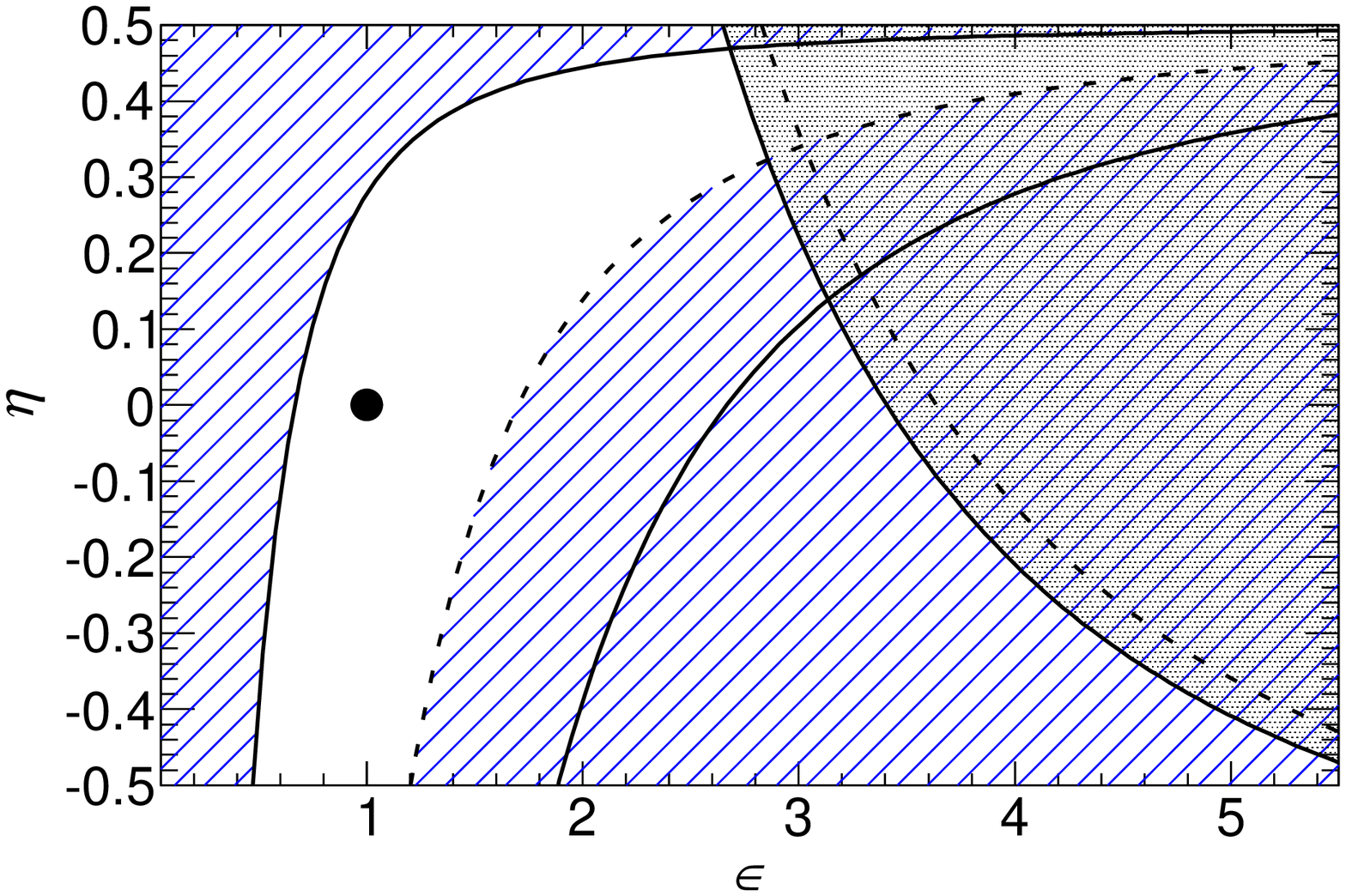} \\
        \caption{\label{fig:KnunuConstraintNew} (color online) The constraints at 90\% CL on $\epsilon$ and $\eta$ of Eq.~\eqref{KnunuEps} for sensitivity to new physics with right-handed currents.  The \kxnunu (diagonal shading) and \ksnunu (grey shading) excluded areas are determined from the upper and lower limits of this \knunu analysis (solid curves) and from the most-stringent upper limits from previous semileptonic-tag analyses \cite{ref:elisaKnunu, ref:carlKnunu} (dashed curves).
The dot shows the expected SM value. }
 \end{figure}

The \knunu decays are also sensitive to the short-distance Wilson coefficients $|C_{L,R}^{\nu}|$ for the left- and right-handed weak currents, respectively.  These couple two quarks to two neutrinos via an effective field theory point interaction \cite{ref:wilson}. Although $|C_{R}^{\nu}|=0$ within the SM, right-handed currents from new physics, such as non-SM \Z penguin couplings, could produce non-zero values.  Using the parameterization from Ref.~\cite{ref:altmann},
\begin{equation}
	\begin{split}
	\label{KnunuEps}
	\epsilon \equiv \frac{\sqrt{|C_L^{\nu}|^2+|C_R^{\nu}|^2}}{|C_{L,{\rm SM}}^{\nu}|}~,~~~
	\eta \equiv \frac{-{\rm Re}(C_L^{\nu}C_R^{\nu*})}{|C_{L}^{\nu}|^2+|C_{R}^{\nu}|^2}~,
	\end{split}
\end{equation}
the \ksnunu upper limits from this search improve the constraints from previous searches on the Wilson-coefficient parameter space, as shown in Fig.~\ref{fig:KnunuConstraintNew}.  The \kxnunu lower limit provides the first upper bound on $\eta$ and lower bound on $\epsilon$.  These constraints are consistent with the expected SM values of $\epsilon=1$ and $\eta=0$. 

\section{\boldmath  Results for \ccnunu}
\label{sec:PhysicsJpsi}

 \begin{figure}[!htbp]
 	\includegraphics[width=2.8in]{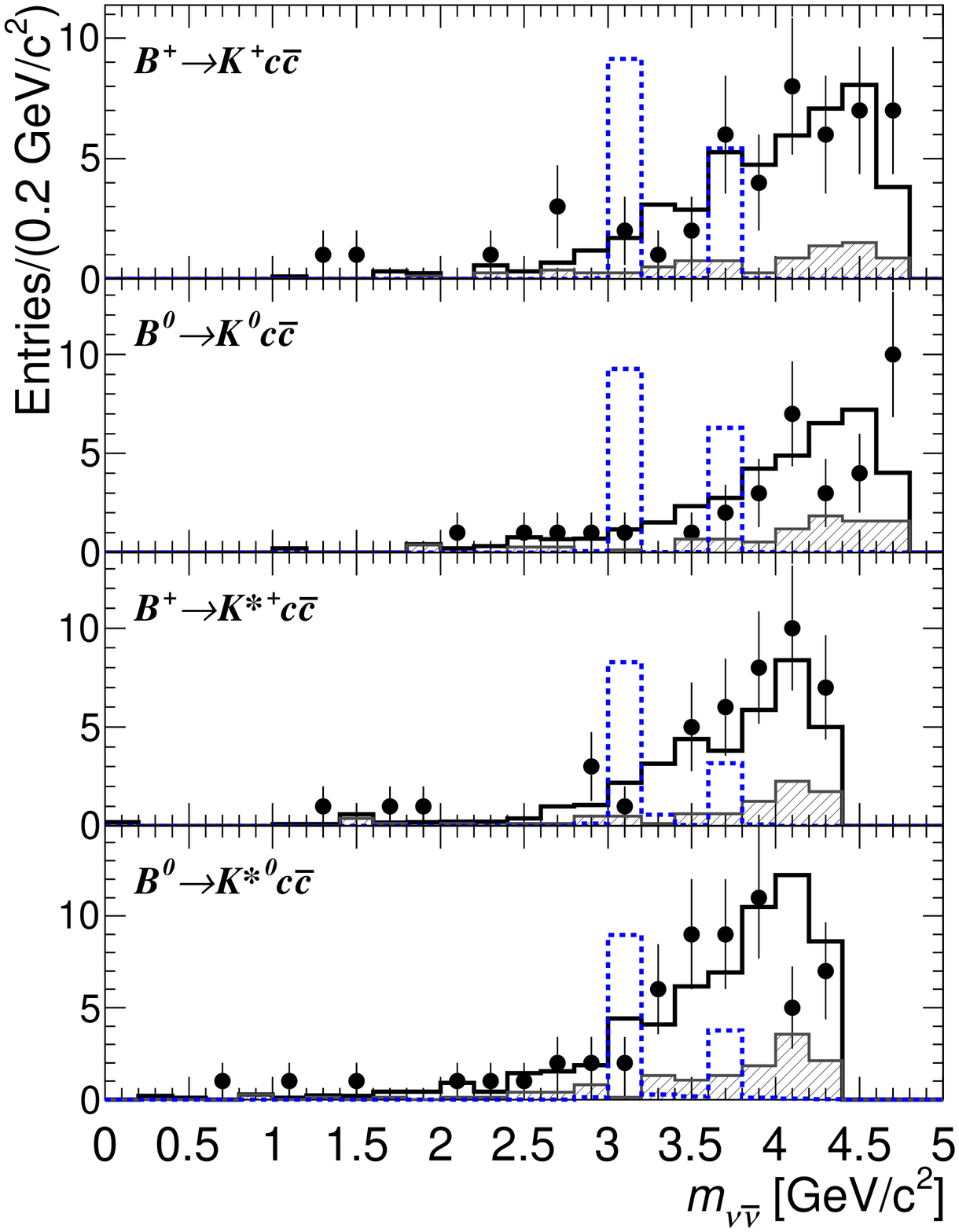} 
 	\caption{\label{fig:recoilMass} (color online) The $m_{\nu\nub} \equiv \sqrt{\sB m_B^2}$ distribution for (from top to bottom) $\Bp\to\Kp\ccbar$, $\Bz\to\Kz\ccbar$, $\Bp\to\Kstarp\ccbar$, and $\Bz\to\Kstarz\ccbar$ events after applying the full signal selection.  The expected combinatorial (shaded) plus \mes-peaking (solid) background contributions are overlaid on the data (points).  The signal MC distributions (dashed) are normalized to $\BR(\ccnunu)$ values of 2\% for the $K^+$ channel, 10\% for the $\Kz$ channel, and 5\% for the $K^{*}$ channels.
} 
 \end{figure}

\begin{table*}[!tbp]
 \caption{Expected \jnunu background yields $\Npeak$ and $\Nbkg$, signal efficiencies \eff, number of observed data events \Nobs, and the resulting branching fraction central value and upper limit at 90\% CL, all within the $m_{\nunub}$ invariant mass region corresponding to the \jpsi mass.  Uncertainties are statistical and systematic, respectively. The \Ncomb yields are calculable as $\Nbkg-\Npeak$. \label{tab:finalNumsJPsi}}
 \begin{ruledtabular}
 \begin{tabular}{lcccccc}
&\multicolumn{6}{c}{\jnunu}\\

Channel & $\Kp$ & $K^0$ & $\Kstarp\to\Kp\piz$ & $\Kstarp\to\KS\pip$ & $\Kstarz\to\Kp\pim$ & $\Kstarz\to\KS\piz$ \\ \hline \hline

\Npeak &$0.4\pm0.2\pm0.0$ & $0.7\pm0.3\pm0.1$ & $0.8\pm0.3\pm0.1$ & $0.4\pm0.2\pm0.0$ & $2.6\pm0.5\pm0.3$ & $0.6\pm0.2\pm0.1$ \\ 
\Nbkg      & $0.5\pm0.2\pm0.0$ & $0.7\pm0.3\pm0.1$ & $0.8\pm0.3\pm0.1$ & $0.8\pm0.3\pm0.0$ & $2.8\pm0.5\pm 0.3$ & $0.6\pm0.2\pm0.1$ \\
\eff $(\times 10^{-8})$ &$95.3\pm4.4\pm5.5$ &$19.3\pm1.0\pm2.1$  &$20.9\pm1.5\pm1.7$ &$12.4\pm0.8\pm1.0$ & $36.2\pm1.9\pm4.0$  & $1.8\pm0.2\pm0.2$\\
\Nobs &1 &0 &1 &0 &0 & 1\\

\multicolumn{2}{l}{$\BR(\jnunu)$} & \multicolumn{4}{c}{$(0.2^{+2.7}_{-0.9}$$^{+0.5}_{-0.4}) \times 10^{-3}$} & \\
\multicolumn{2}{l}{Limit} & \multicolumn{4}{c}{$<3.9 \times 10^{-3}$} &\\

 \end{tabular}
 \end{ruledtabular}
\end{table*}

\begin{table*}[!tbp]
         \caption{Expected \psinunu background yields $\Npeak$ and $\Nbkg$, signal efficiencies \eff, number of observed data events \Nobs, and the resulting branching fraction central value and upper limit at 90\% CL, all within the $m_{\nunub}$ invariant mass region corresponding to the \psitwos mass.  Uncertainties are statistical and systematic, respectively. The \Ncomb yields are calculable as $\Nbkg-\Npeak$. \label{tab:finalNumsPsi2S}}
         \begin{ruledtabular}
         \begin{tabular}{lcccccc}
&\multicolumn{6}{c}{\psinunu}\\

Channel & $\Kp$ & $K^0$ & $\Kstarp\to\Kp\piz$ & $\Kstarp\to\KS\pip$ & $\Kstarz\to\Kp\pim$ & $\Kstarz\to\KS\piz$ \\ 
\hline \hline

\Npeak & $1.4\pm0.4\pm0.1$ & $0.6\pm0.3\pm0.1$ & $1.4\pm0.4\pm0.1$ & $1.0\pm0.3\pm0.1$ & $3.5\pm0.7\pm0.3$ & $0.6\pm0.2\pm0.1$ \\ 
\Nbkg & $1.6\pm0.4\pm0.1$ & $0.7\pm0.3\pm0.1$ & $1.4\pm0.4\pm0.1$ & $1.5\pm0.4\pm0.1$ & $3.9\pm0.7\pm0.3$ & $0.6\pm0.2\pm0.1$  \\
\eff $(\times 10^{-8})$  & $57.2\pm3.5\pm3.3$& $13.1\pm1.2\pm1.4$ & $8.1\pm1.7\pm0.7$&$4.9\pm1.1\pm0.4$ &$14.2\pm1.2\pm 1.6$  & $0.6\pm0.1\pm0.1$\\
\Nobs     & 3 & 1 & 1 & 3 & 5 &1 \\

\multicolumn{2}{l}{$\BR(\psinunu)$} & \multicolumn{4}{c}{$(5.6^{+7.4}_{-4.6}$$^{+1.6}_{-1.4})\times 10^{-3}$} &\\
\multicolumn{2}{l}{Limit}  & \multicolumn{4}{c}{$<15.5\times 10^{-3}$} &\\

         \end{tabular}
         \end{ruledtabular}
\end{table*}

In the search for \ccnunu, Fig.~\ref{fig:recoilMass} shows the $m_{\nu\nub}$ distribution of the observed data yields, expected background contributions, and SM signal distributions.  Tables \ref{tab:finalNumsJPsi} and \ref{tab:finalNumsPsi2S} summarize the background contribution values and signal efficiencies within the \jpsi and \psitwos invariant mass regions.  The tables also report the combined branching fraction central values and the branching fraction upper limits at 90\% CL for \jnunu and \psinunu.  The signal efficiencies account for the $\B\to\kaon^{(*)}\jpsi$ and $\B\to\kaon^{(*)}\psitwos$ branching fractions and their errors, which are taken from Ref.~\cite{ref:pdg}. The data yield is consistent with zero observed \ccnunu signal events in all channels.

The combined upper limits for the charmonium branching fraction values are determined to be 
\begin{equation}
	\label{ratios}
	\begin{split}
 	&\frac{\BR(\jnunu)}{\BR(\jpsi\to\epem)}< 6.6\times 10^{-2}~{\rm and}  \\
        &\frac{\BR(\psinunu)}{\BR(\psitwos\to\epem)}< 2.0,
	\end{split}
\end{equation}
where $\BR(\jpsi\to\epem)$ and $\BR(\psitwos\to\epem)$ are taken from Ref.~\cite{ref:pdg}.  With the addition of a new-physics $U$ boson, these ratios would be proportional to $|f_{cV} c_{\chi, \varphi}|$, where $c_{\chi, \varphi}$ and $f_{cV}$ are the $U$ couplings to the LDM particles $\chi$ or $\varphi$ and to the $c$-quark respectively \cite{ref:fayet}. The \jpsi decay ratio yields upper limits at 90\% CL of $|f_{cV} c_{\chi, \varphi}|<(3.0, 2.1, 1.5) \times 10^{-2} $ for spin-0, Majorana, and Dirac LDM particles respectively.  These limits are comparable with those obtained by BES for \jnunu, in $\psitwos\to\pip\pim\jpsi$ \cite{ref:BES}.

\section{Summary}
\label{sec:Summary}

In conclusion, we have searched for the decays \kxnunu and \ksnunu, as well as \jnunu and \psinunu via $B\to \KorKstar\jpsi$ and $B\to \KorKstar\psitwos$, recoiling from a hadronically reconstructed $B$ meson within a data sample of $471\times10^6$ \BB pairs.  We observe no significant signal in any of the channels and obtain upper limits at the 90\% CL of $\BR(\kxnunu)<3.2\times10^{-5}$, $\BR(\ksnunu)<7.9\times10^{-5}$, $\BR(\jnunu)<3.9\times 10^{-3}$, and $\BR(\psinunu)<15.5\times 10^{-3}$.  The branching fraction central values and upper limits are consistent with SM predictions.  We report \knunu branching fraction limits in Tables~\ref{tab:finalNums2} and \ref{tab:finalNums}, and \ccnunu branching fraction limits in Tables~\ref{tab:finalNumsJPsi} and \ref{tab:finalNumsPsi2S}. These results include the first lower limit in the \kpnunu decay channel, the most stringent published upper limits using the hadronic-tag reconstruction technique in the \kznunu, \kspnunu, and \ksznunu channels, and the first upper limit for \psitwos\to\nunub.  We also present partial branching fraction values for \knunu over the full \sB spectrum in Fig.~\ref{fig:sBULs} in order to enable additional tests of  new-physics models.   

\section{Acknowledgments}
\label{sec:Acknowledgments}

We are grateful for the 
extraordinary contributions of our \pep2\ colleagues in
achieving the excellent luminosity and machine conditions
that have made this work possible.
The success of this project also relies critically on the 
expertise and dedication of the computing organizations that 
support \babar.
The collaborating institutions wish to thank 
SLAC for its support and the kind hospitality extended to them. 
This work is supported by the
US Department of Energy
and National Science Foundation, the
Natural Sciences and Engineering Research Council (Canada),
the Commissariat \`a l'Energie Atomique and
Institut National de Physique Nucl\'eaire et de Physique des Particules
(France), the
Bundesministerium f\"ur Bildung und Forschung and
Deutsche Forschungsgemeinschaft
(Germany), the
Istituto Nazionale di Fisica Nucleare (Italy),
the Foundation for Fundamental Research on Matter (The Netherlands),
the Research Council of Norway, the
Ministry of Education and Science of the Russian Federation, 
Ministerio de Econom\'{\i}a y Competitividad (Spain), and the
Science and Technology Facilities Council (United Kingdom).
Individuals have received support from 
the Marie-Curie IEF program (European Union) and the A. P. Sloan Foundation (USA).


\begin{thebibliography}{99}
\bibitem{ref:altmann} W.\ Altmannshofer, A.\ J.\ Buras, D.\ M.\ Straub, and M.\ Wick, JHEP {\bf0904}, 022 (2009).
\bibitem{ref:chang} L.\ N.\ Chang, O.\ Lebedev, and J.\ N.\ Ng, Phys.\ Lett. B {\bf441}, 419 (1998).
\bibitem{ref:buchalla} G.\ Buchalla, G.\ Hiller, and G.\ Isidori, Phys.\ Rev. D {\bf63}, 014015 (2000).
\bibitem{ref:LYY} X.\ -Q.\ Li, Y.\ -D.\  Yang, and  X.\ -B.\ Yuan, JHEP  {\bf1203}, 018 (2012).
\bibitem{ref:leptoZ} J.\ H.\ Jeon,   C.\ S.\ Kim,  J.\ Lee, and C.\ Yu, Phys.\ Lett.  B {\bf636}, 270 (2006).
\bibitem{ref:bird} C.\ Bird, R.\ V.\ Kowalewski, and M.\ Pospelov, Mod.\ Phys.\ Lett  A {\bf21}, 457 (2006).
\bibitem{ref:mckeen} D.\ McKeen, Phys.\ Rev.  D{\bf 79}, 114001 (2009).
\bibitem{ref:darkBosons} H.\ Davoudiasl, H.- S.\ Lee, and W.\ J.\ Marciano, Phys. Rev. D {\bf85}, 115019 (2012).
\bibitem{ref:unparticle} T.\ M.\ Aliev,  A.\ S.\ Cornell, and N.\ Gaur, JHEP  {\bf0707}, 072 (2007).
\bibitem{ref:FCNCportals} J.\ F.\ Kamenik and C.\ Smith, JHEP {\bf1203}, 090 (2012).
\bibitem{ref:ColangeloKnunu} P.\ Colangelo, F.\ De Fazio, R.\ Ferrandes, and T.\ N.\ Pham, Phys.\ Rev.  D {\bf73}, 115006 (2006).
\bibitem{ref:McElrath} B.\ McElrath, Phys.\ Rev. D {\bf72}, 103508 (2005).
\bibitem{ref:fayet} P.\ Fayet, Phys.\ Rev. D {\bf74}, 054034 (2006).
\bibitem{ref:ccNote} {The use of charge conjugate processes is implied throughout this article.}
\bibitem{ref:carlKnunu} P.\ del Amo Sanchez {\em et al.} [\babar\ Collaboration], Phys.\ Rev. D {\bf82}, 112002 (2010).
\bibitem{ref:elisaKnunu} B.\ Aubert {\em et al.} [\babar\ Collaboration], Phys.\ Rev. D {\bf78}, 072007 (2008).
\bibitem{ref:jackKnunu} B.\ Aubert {\em et al.} [\babar\ Collaboration], Phys.\ Rev.\ Lett. {\bf94}, 101801 (2005).
\bibitem{ref:newBelle} O.\ Lutz {\em et al.} [BELLE Collaboration], arXiv:1303.3719. 
\bibitem{ref:belleKnunu} K.- F.\ Chen {\em et al.} [BELLE Collaboration], Phys.\ Rev.\ Lett. {\bf99}, 221802 (2007).

\bibitem{ref:BES} M.\ Ablikim {\em et al.} [BES Collaboration], Phys.\ Rev.\ Lett.\ {\bf100}, 192001 (2008).
\bibitem{ref:lumi} J.\ P.\ Lees {\it et al.} [\babar\ Collaboration], arXiv:1301.2703 [Nucl.\ Instr.\ Meth. A (in press)].
\bibitem{ref:babar} B.\ Aubert {\em et al.} [\babar\ Collaboration], Nucl.\ Instr.\ Meth. A {\bf479}, 1 (2002).
\bibitem{ref:NIMU} B.\ Aubert {\em et al.} [\babar\ Collaboration], arXiv:1305.3560 [Nucl.\ Instr.\ Meth. A (in press)].
\bibitem{ref:geant4} S.\ Agostinelli {\em et al.} [Geant4 Collaboration], Nucl.\ Instr.\ Meth. A {\bf506}, 250 (2003).
\bibitem{ref:jpsikamp} B.\ Aubert {\em et al.} [\babar\ Collaboration], Phys.\ Rev.\ D {\bf76}, 031102 (2007).
\bibitem{ref:Dtaunu} J.\ P.\ Lees {\em et al.} [\babar\ Collaboration], Phys.\ Rev.\ Lett. {\bf109}, 101802 (2012).
\bibitem{ref:punzi} G.\ Punzi, eConf {\bf C030908}, MODT002 (2003); arXiv:physics/0308063.
\bibitem{ref:thrust} E.\ Farhi, Phys.\ Rev.\ Lett. {\bf39}, 1587 (1977).
\bibitem{ref:foxWolf} G.\ C.\ Fox and S.\ Wolfram, Phys.\ Rev.\ Lett. {\bf41}, 1581 (1978).
\bibitem{ref:pdg} K.\ Nakamura {\em et al.} [Particle Data Group], J. Phys. G {\bf37}, 075021 (2010), and 2011 partial update for the 2012 edition.
\bibitem{ref:bartsch} M.\ Bartsch, M.\ Beylich, G.\ Buchalla, and D.- N.\ Gao, JHEP {\bf0911}, 011 (2009).
\bibitem{ref:barlow} R.\ Barlow, Comput.\ Phys.\ Commun. {\bf149}, 97 (2002).
\bibitem{ref:wilson} G.\ Buchalla, A.\ J.\ Buras, and M.\ E.\ Lautenbacher, Rev.\ Mod.\ Phys. {\bf68}, 1125 (1996).


\end{thebibliography}
\end{document}